\input epsf

\def\INSERTCAP#1#2{\vbox{%
{\narrower\noindent%
\multiply\baselineskip by 3%
\divide\baselineskip by 4%
{\rm Table #1. }{\sl #2 \medskip}}
}}
\input lanlmac
\def\href#1#2{{#2}}
%
\def\eqn#1#2{\xdef #1{(\secsym\the\meqno)}\writedef{#1\leftbracket#1}%
\global\advance\meqno by1$$#2\leqno#1\eqlabeL#1$$}
%
\def\subsec#1{\global\advance\subsecno by1\message{(\secsym\the\subsecno. #1)}
\ifnum\lastpenalty>9000\else\bigbreak\fi
\noindent{\it\the\secno$\cdot$\the\subsecno. #1}\writetoca{\string\quad 
{\secsym\the\subsecno.} {#1}}\par\nobreak\medskip\nobreak}
\font\smallheadfont=cmcsc10
\def\HF{\smallheadfont}
%
%
\def\OMIT#1{}

\def\eg{{\it e.g.,\/\ }}

\def\spur{\raise.15ex\hbox{/}\kern-.57em }

\def\CO{{\cal O}}
\def\ccdot{\hbox{\kern-.1em$\cdot$\kern-.1em}}
\def\frac#1#2{{#1\over#2}}

\def\larr#1{\raise1.5ex\hbox{$\leftarrow$}\mkern-16.5mu #1}

%
%

%
%

\def\lsl{\spur {\kern0.1em l}}

%

%

\def\np#1#2#3{\NP{\bf #1} (#2) #3}
\def\pl#1#2#3{\PL{\bf #1} (#2) #3}
\def\prl#1#2#3{\PRL{\bf #1} (#2) #3}

\def\physrev#1#2#3{\PR{\bf #1} (#2) #3}

\def\NP{{\it Nucl.\ Phys.\ }}
\def\PL{{\it Phys.\ Lett.\ }}
\def\PR{{\it Phys.\ Rev.\ }}

\def\PRL{{\it Phys.\ Rev.\ Lett.\ }}

%
%
%
\catcode`\@=11 
\global\newcount\exerno \global\exerno=0
\def\exercise#1{\begingroup\global\advance\exerno by1%
\medbreak
\baselineskip=10pt plus 1pt minus 1pt
\parskip=0pt
\hrule\smallskip\noindent{\sl Exercise \the\secno.\the\exerno \/}
{\ninepoint #1}\smallskip\hrule\medbreak\endgroup}
%
%
\def\newsec#1{\global\advance\secno by1\message{(\the\secno. #1)}
\global\exerno=0%
\global\subsecno=0\eqnres@t\noindent{\bf\the\secno. #1}
\writetoca{{\secsym} {#1}}\par\nobreak\medskip\nobreak}
\def\appendix#1#2{\global\exerno=0%
\global\meqno=1\global\subsecno=0\xdef\secsym{\hbox{#1.}}
\bigbreak\bigskip\noindent{\bf Appendix #1. #2}\message{(#1. #2)}
\writetoca{Appendix {#1.} {#2}}\par\nobreak\medskip\nobreak}
\catcode`\@=12 
%
%
%
%

%
\relax


%
\def\w{{\omega}}

\Title{\vbox{\hbox{UCSD/PTH 95-15}}}{\vbox{%
\centerline{Model-Independent Semileptonic Form Factors} 
\centerline{Using Dispersion Relations$^{\S}$}}}
\centerline{C. Glenn Boyd\footnote{}{\hskip-2.5ex$^{\S}$\hskip1ex{}Talk
presented by B.G. at the 6-th International Symposium on Heavy Flavour
Physics, Pisa, Italy, 6--10 June, 1995}\footnote{$^{\ast}$}{gboyd@ucsd.edu},
Benjam\'\i n Grinstein\footnote{$^{\dagger}$}{bgrinstein@ucsd.edu} and
Richard F. Lebed\footnote{$^{\ddagger}$}{rlebed@ucsd.edu}}
\bigskip\centerline{Department of Physics}
\centerline{University of California, San Diego}
\centerline{La Jolla, California 92093-0319}

\vskip .3in
 	We present a method for parametrizing heavy meson semileptonic
form factors using dispersion relations, and from it produce a
two-parameter description of the $B \to B$ elastic form factor.  We
use heavy quark symmetry to relate this function to $\bar B\to D^* l
\bar \nu$ form factors, and extract $|V_{cb}|=0.0355^{+0.0029}_{-0.0025}$
from experimental data with a least squares fit.  Our method
eliminates model-dependent uncertainties inherent in choosing a
parametrization for the extrapolation of the differential decay rate
to threshold. 
%
\Date{June 1995} 
\centerline{\bf Model-Independent Semileptonic Form Factors 
Using Dispersion Relations}
\centerline{C. Glenn Boyd,
Benjam\'\i n Grinstein and
Richard F. Lebed}
\centerline{Department of Physics, University of California, San Diego}
\centerline{La Jolla, California 92093-0319}

\bigskip
\leftskip=1.0cm
\rightskip=1.0cm{\noindent
     	We present a method for parametrizing heavy meson semileptonic
form factors using dispersion relations, and from it produce a
two-parameter description of the $B \to B$ elastic form factor.  We
use heavy quark symmetry to relate this function to $\bar B\to D^* l
\bar \nu$ form factors, and extract $|V_{cb}|=0.0355^{+0.0029}_{-0.0025}$
from experimental data with a least squares fit.  Our method
eliminates model-dependent uncertainties inherent in choosing a
parametrization for the extrapolation of the differential decay rate
to threshold. 
}

\leftskip=0cm
\rightskip=0cm

\newsec{ Introduction}

  	A non-perturbative, model-independent description of QCD form
factors is a desirable ingredient for the extraction of
Cabibbo-Kobayashi-Maskawa parameters from exclusive meson decays.
Progress towards this goal has been realized by the development of
heavy quark symmetry\ref\hqs{{\HF N. Isgur and M. B. Wise},
\pl{B232}{1989}{113} and {\bf B237}~(1990)~527\semi {\HF E.  Eichten
and B. Hill}, \pl{B234}{1990}{511}\semi {\HF M. B.  Voloshin and M. A.
Shifman},{\it Yad.\ Fiz.}\ {\bf 47} (1988) 801  [{\it Sov.\ J.\ Nucl.\
Phys.}\ {\bf 47} (1988) 511].}, which relates and normalizes the
$\bar B \to D^* l \bar{\nu}$ and $\bar B \to D l
\bar{\nu}$ form factors in the context of a ${1\over M}$ expansion,
where $M$ is the heavy quark mass.  Previous
talks\ref\elsewherepisa{See, \eg talks by {\HF V. Sharma, I. Shipsey}
and {\HF M. Schmitt}, these proceedings.} have described how this
normalization is used%
\nref\cleo{{\HF B. Barish {\it et al.} (CLEO Collaboration)},
\physrev{D51}{1995}{1014}.}\nref\arg{{\HF H. Albrecht {\it et al.}
(ARGUS Collaboration)}, {\it Z.~Phys.}\ {\bf C57} (1993)
533.}\nref\aleph{{\HF I.~Scott (ALEPH Collaboration)}, to appear in
{\it Proceedings of the 27th International Conference on High Energy
Physics, Glasgow, Scotland, July 1994.}}\refs{\cleo{--}\aleph}\
to extract the value of the CKM parameter $|V_{cb}|$ by extrapolating
the measured form factor to zero recoil, where the normalization is
predicted.

  	This form factor extrapolation, necessary because the rate
vanishes at zero recoil, introduces an uncertainty in the value of
$|V_{cb}|$ due to the choice of parametrization.  Estimates of this
uncertainty obtained by varying parametrizations suffer the same
ambiguity. This ambiguity could be eliminated if one had a
non-perturbative, model-independent characterization of the form factor
in terms of a small number of parameters.

\nref\usone{{\HF C. G. Boyd, B. Grinstein and R. F. Lebed},
\pl{B353}{1995}{306}  \hfil\break 
\href{http://xxx.lanl.gov/abs/hep-ph/9504235}{[hep-ph/9504235]}.}
\nref\ustwo{{\HF C. G. Boyd, B. Grinstein and R. F. Lebed},
\prl{74}{1995}{4603}\hfil\break 
\href{http://xxx.lanl.gov/abs/hep-ph/9412324}{[hep-ph/9412324]}.}

In this talk we describe recent work\refs{\usone{--}\ustwo}\ in which
we use dispersion relations to derive such a characterization and
apply it towards the extraction of $|V_{cb}|$.  Our characterization
of the $B$ elastic $b$-number form factor uses two parameters and has
1\% accuracy over the entire physical range relevant to the extraction
of $|V_{cb}|$. What is important is not that we effectively find a
quadratic parametrization of the Isgur-Wise function, but rather that
we have determined the associated uncertainty.

	The outline of the talk is as follows. We first review the
expected achievable accuracy in the extraction of $|V_{cb}|$ from
semi-inclusive semi-leptonic decays in Sec.\ 2. In Sec.\ 3 we
review a well-known method\ref\hist{{\HF N. N.  Meiman}, {\it Sov.\
Phys.\ JETP} {\bf 17} (1963) 830\semi {\HF S.~Okubo and I.~Fushih},
\physrev{\bf D4}{1971}{2020}\semi {\HF V. Singh and A. K.  Raina},
{\it Fortschritte der Physik} {\bf 27} (1979) 561\semi {\HF
C. Bourrely, B. Machet, and E. de Rafael},
\np{B189}{1981}{157}.}\  for using QCD dispersion relations and
analyticity to place constraints on hadronic form factors, and show
that a parametrization of the $B$ elastic form factor $F(q^2)$ in
terms of two parameters is accurate to 1\% over the relevant kinematic
region.  In Sec.\ 4 we use heavy quark symmetry to relate $F(q^2)$ to the
Isgur-Wise function, which describes the form factors for $\bar B \to
D^* l \bar{\nu}$ in the infinite quark mass limit. We make a least
squares fit to CLEO\cleo, ARGUS\arg, and ALEPH\aleph\ data using
$|V_{cb}|$ and our two basis function parameters as variables, and
present our results.  Reliability of the method is discussed in Sec.\
5, implications for $|V_{ub}|$ and $B\to K^*\gamma$ are discussed in
Sec.\ 6, and a summary in Sec.\ 7.

\def\INC{\Gamma(\bar B\to X_c\ell\bar\nu_\ell)}

\newsec{Digression: $|V_{cb}|$ from inclusive semi-leptonic decays.}
Several competing methods for the determination of $|V_{cb}|$ were
described in previous talks\elsewherepisa. We would like to discuss
briefly the theoretical limitations in the determination of $|V_{cb}|$
from inclusive semi-leptonic decays. The interpretation of the
measurement of the inclusive semileptonic rate $\INC$ relies on our
ability to calculate the rate from first principles. Using a Heavy
Quark expansion one can show two things\ref\chay{{\HF J. Chay,
H. Georgi and B. Grinstein,} \pl{B247}{1990}{399}.}:
\item{(i)}The leading term (in $1/m_b$) is given by  the parton
decay rate  $\Gamma(b\to c\ell\bar\nu_\ell)$
\item{(ii)}There are no first order (in $1/m_b$) corrections to the
previous statement.

\nref\wiserev{{\HF M. B. Wise}, {\sl Caltech Report} No.\ CALT-68-1963
\href{http://xxx.lanl.gov/abs/hep-ph/9411264}{[hep-ph/9411264]}.}
\nref\shifrev{{\HF  M. Shifman}, {\sl Theor. Phys. Inst. Report} No.\
TPI-MINN-94/31-T
\href{http://xxx.lanl.gov/abs/hep-ph/9409358}{[hep-ph/9409358]}.} 

In a $1/m_b$ expansion one may write\refs{\wiserev,\shifrev}
\eqn\ttlrate{\INC=\Gamma_0[A(x)\eta+B_K(x)K_b+B_G(x)G_b+\CO(1/m_b^3)].
}
There is a lot of notation to explain here. First, $\Gamma_0A(x)$ is
just the parton decay rate, with
$\Gamma_0=G_F^2m_b^5|V_{cb}|^2/192\pi^3$, $x=m_c^2/m_b^2$ and
$A(x)=1-8x+8x^3-x^4-12x^2\ln x$ is a kinematic factor from the phase
space integral. $\eta$ is a correction factor\wiserev\ from
perturbative QCD,
\eqn\etagiven{
\eta=1-\left({\alpha\over\pi}\right)\Delta^{(1)}(x)-\left({\alpha\over\pi}\right)^2
(\beta_0\Delta^{(2)}_{\beta_0}(x)+\cdots)+\CO(\alpha^3).
}
The one loop function $\Delta^{(1)}(x)$ can be computed analytically and
equals 1.7 at $x=(0.3)^2$ (2.8 at $x=0$). The full two-loop
computation is not available, but the part proportional to $\beta_0$, the
one-loop beta function, has been computed:  $\Delta_{\beta_0}^{(2)}=1.7$
at   $x=(0.3)^2$ (3.2 at $x=0$).
 
The next two terms are the $1/m_b^2$ corrections, with
\eqn\kandg{K_b={1\over2m_B}\vev{B|\bar b{D^\mu
D_\mu\over2m_b^2}b|B}\quad{\rm and}\quad
G_b={1\over2m_B}\vev{B|\bar b{g_s\sigma^{\mu\nu}G_{\mu\nu}\over4m_b^2}b|B}
}
and the corresponding kinematic factors $B_K(x)=-A(x)$ and $B_G(x)=3 -
8x+24x^2-24x^3+5x^4+12x^2\ln x$. 

\def\IBUL{\item{$\bullet$}}
There are three main sources of thoretical uncertainties:
\IBUL{\it Quark masses} enter the rate in the combination
$m_b^5 A(x)$. This is less sensitive to uncertainties in $m_b$ than
$m_b^5$ if $m_c$ is fixed by
\eqn\mcgiven{
m_B-m_D=m_b-m_c+m_b(K_b+G_b)-m_c(K_c+G_c)
}
For a fixed guess of $K_{b,c}$ and $G_{b,c}$ Wise\wiserev\
(Shifman\shifrev) finds that $\Delta|V_{cb}|/|V_{cb}|=10\%$ (3.2\%)
for $\Delta m_b=0.5$~GeV (0.2~GeV).
\IBUL {\it Perturbative QCD corrections} are still not fully computed at
two loops. This is not a limitation in principle, except in the case
that the expansion does not converge quickly.\foot{The expansion is
asymptotic, so it does not converge. The asymptotic expansion has an
accuracy equal to the size of the term at which the expansion is
truncated.} With $x=(0.3)^2$, $\beta_0=9$ and $\alpha_s=0.2$ the three
terms in \etagiven\ are\wiserev\ $\eta=1-0.11-0.06$
($\eta=1-0.15-0.11$ for $x=0$), which appears only marginally
convergent.
\IBUL{\it Non-perturbative corrections} come in through the parameters $K$
and $G$. The latter is well know as it determines the  $B^*-B$ mass
difference. Shifman\shifrev\ estimates an uncertainty of
$\Delta|V_{cb}|/|V_{cb}|=5.6\%$ from $K_b=0.024\pm0.009$.

I close with a digression on a method related but not the same as the
heavy quark expansion. One may argue that $\INC=\Gamma(b\to
c\ell\bar\nu_\ell)$ in the limit $m_b-m_c\ll\Lambda_{\rm QCD}\ll
m_{c,b}$. The argument is straightforward. In this limit only the $D$
and $D^*$ resonances are kinematically allowed, so they saturate the
semi-inclusive rate. The rates into these resonances are determined
effectively by the Isgur-Wise function at $v\cdot v'=1$, where it is
normalized to unity. The resulting rate is precisely $\Gamma(b\to
c\ell\bar\nu_\ell)$, provided one does not differentiate between quark
and meson masses. 

In the limit, the rate for $\Gamma(\bar B\to
X\ell\bar\nu_\ell)\sim(m_B-m_X)^5$, for $X=D,D^*$. This can be used to
`explain' the lower sensitivity of $\INC$ to $m_b$ when $m_c$ is fixed
in terms of $m_b$ using the known value of $m_B-m_D$. But the real
question is whether this could be the basis for a systematic
expansion. Note, however, that
\eqna\SVdisaster
$$\eqalignno{
(m_B-m_{D})^5 &=(m_b-m_c)^5(1+K+G)^5 +\cdots &\SVdisaster a\cr
(m_B-m_{D^*})^5&=(m_b-m_c)^5(1+K-G)^5 
\left(1+10G{m_b+m_c\over m_b-m_c}\right) +\cdots  &\SVdisaster b \cr
}
$$
These correction factors will enter  the relation between
inclusive and partonic widths. The expansion parameter in
Eq.~\SVdisaster b\ is poor: $G{m_b+m_c\over m_b-m_c}\sim 
{\Lambda_{\rm QCD}\over m_b-m_c}{\Lambda_{\rm QCD}\over m_b}$.

\nref\drtone{{\HF E.  de
Rafael and J. Taron}, \pl{B282}{1992}{215}.}%
\nref\spoiler{ {\HF E. Carlson, J. Milana, N. Isgur, T.
Mannel, and W. Roberts}, \pl{B299}{1993}{133}\semi
{\HF A. Falk, M. Luke, and M. Wise}, \pl{B299}{1993}{123}\semi 
{\HF B. Grinstein and P. Mende}, \pl{B299}{1993}{127}\semi 
{\HF J. K\"orner and D. Pirjol}, \pl{B301}{1993}{257}.}%
\nref\drttwo{{\HF E. de Rafael and J. Taron}, 
\physrev{50}{1994}{373}.}%

\newsec{The Analyticity Constraints$^2$}
     
      Of primary interest is the form factor $F$, defined
by\footnote{\null}{\hskip-2.5ex$^2$\hskip1ex This section based on
refs.~\refs{\hist,\drtone{--}\drttwo}}
\eqn\ffsdefd{ 
\vev{B(p')| V_\mu | B(p)} = F(q^2)(p+p')_\mu ,
}
where $V_\mu=\bar b \gamma_\mu b$, and $q^2 = (p-p')^2$. Crossing
symmetry states that the form factor $F$ is a function of a complex
variable, $q^2$, which gives the matrix element in Eq.~\ffsdefd\ for
real negative $q^2$, and the matrix element $\vev{0|V_\mu|B\bar B}$
for real $q^2\ge4m_B^2$.

Consider the two-point function
\eqn\twopntfnctn{
i \int d^4\!x e^{iqx}\vev{{\rm T} V_\mu(x) V^\dagger_\nu(0)}
=(q_\mu q_\nu-q^2g_{\mu\nu})\Pi(q^2) 
}
In QCD the structure functions satisfy
a once-subtracted dispersion relation:
\eqn\dsptnrltn{
\left.{{\partial\Pi}
\over{\partial q^2}}\right|_{q^2=-Q^2}=
{1\over\pi}\int_0^\infty dt \, {{{\rm Im}\,\Pi(t)}\over{(t+Q^2)^2}}
.}
The absorptive part ${\rm Im}\,\Pi(q^2)$ is obtained by inserting real
states between the two currents on the right-hand side of
Eq.~\twopntfnctn. This is a
sum of positive definite terms, so one can obtain strict inequalities
by concentrating on the term with intermediate states of $B\bar B$
pairs.  For $Q^2$ far from the resonance region ($Q^2+4m_b^2\gg
m_b\Lambda_{\rm QCD}$) the two-point function
can be computed reliably from perturbative QCD. We set $Q^2=0$ which for
large $b$ quark mass is far from resonances. One thus obtains
an inequality of the form
\eqn\onegf{
\int_{4M_B^2}^\infty dt\, k(t) |F|^2 \le 1,
} 
where the function $k(t)$ is the ratio of the kinematic factor on the
right hand side of Eq.~\dsptnrltn\ to the QCD calculation of the left
hand side.

	A key ingredient in this approach is the transformation that
maps the complex $q^2$ plane onto the unit disc $|z| \leq 1$:
\eqn\zdef{
\sqrt{1-{q^2 \over 4M_B^2}} = {{1+z}\over{1-z}}~.
}
In terms of the angular variable $e^{i \theta} \equiv z$, the
once-subtracted QCD dispersion relation may be written as
\eqn\disp{
{1\over2\pi} \int_0^{2\pi} d\theta\, |\phi(e^{i \theta}) F
(e^{i \theta})|^2 \leq {1\over\pi}~,
}
where the weighing function $\phi(e^{i
\theta})=k(t(\theta))dt/d\theta$\drtone:
\eqn\phidef{
\phi (z) = {1 \over 16}\sqrt{{5 n_f} \over 6 \rho} (1 +z)^2 \sqrt{1-z} .
}
\nref\rein{{\HF L. J. Reinders, H. R. Rubinstein, and S. Yazaki},
\np{B186}{1981}{109}\semi
{\HF M. A. Shifman, A. I. Vainshtein, M. B. Voloshin, and
V. I.  Zakharov}, \pl{B77}{1978}{80}\semi 
{\HF M. A. Shifman, A.  I. Vainshtein, and V. I. Zakharov}, 
\np{B147}{1979}{385}.}%
Here $n_f$ is the number of light flavors for
which $SU(n_f)$ flavor symmetry is valid; we take $n_f =2$.
Perturbative corrections to the dispersion relation are incorporated
in $\rho$, which has been computed\rein\ to
$\CO(\alpha_s)$, $\rho = 1 + 0.73
\alpha_s(m_b) \approx 1.20$.

Note that $\phi(z)$ is analytic in $|z|<1$, while poles of $F$ inside
the unit disc originate from resonances below threshold and cannot be
ignored\spoiler. In this talk we only consider the effects of the
resonances $\Upsilon_{1,2,3}$. Although cuts below threshold should be
considered, they are expected to have smaller physical effects.  A
simple but effective trick\drttwo\ eliminates the poles with no
reference to the size of their residues but rather only their
positions ({\it i.e.}, masses).  The function
\eqn\pp{
P(z) \equiv {(z -z_1)(z -z_2)(z -z_3) \over (1 - \bar z_1 z)(1 - \bar
        z_2 z)(1 - \bar z_3 z)}, 
} 
where the $z_i$ correspond to the values $q^2 = M^2_{\Upsilon_i}$, is
analytic in $|z|\le1$ and satisfies $|P(z)|=1$ for $|z|=1$. The
function $P(z)\phi(z)F(z)$ is analytic on the unit disk, and obeys
\eqn\Idef{
{1\over2\pi} \int_0^{2\pi} d\theta\, 
|P(e^{i \theta})\phi(e^{i \theta}) F
(e^{i \theta})|^2 \leq {1\over\pi}~. 
}
It follows that the QCD form factor may therefore be written as
\eqna\basisfn
$$
F(z) = {1\over P(z) \phi(z)} \sum_{n=0}^{\infty} a_n z^{n} 
\qquad{\rm with}\qquad
\sum_{n=0}^{\infty} 
|a_n|^2 \le {1 \over \pi}\leqno\basisfn {a,b}
$$
Since $B$-number is conserved, $F(0)=1$, so that $a_0=P(0)\phi(0)$.

In the next section we will use heavy quark symmetries to relate $F$
to the form factors for $\bar B\to D^* l\bar\nu$, where the physical
kinematic range is $0 < z < 0.056$. Therefore, the form factor in
\basisfn a\ converges quickly over the physical region. Note that for
this it is crucial that the coefficients $a_n$ be bounded as in
\basisfn b.  Retaining only $a_1$ and $a_2$ in Eq.~\basisfn a, we have
\eqn\parametriztn{
F(z)={1\over P(z)\phi(z)}[P(0)\phi(0)+a_1z+a_2z^2]
}
with a maximum relative error of $\sim{\sqrt{1/\pi}(0.056)^3 / P(0)
\phi(0)} \approx 0.01 $.

\newsec{Extraction of $|V_{cb}|$}

\subsec{Heavy Quark Symmetry Relations}
	In the infinite $b$ and $c$ quark mass limit all the form
factors for $\bar B \to D l \bar \nu$ and $\bar B \to D^* l \bar \nu$
are given by one universal ``Isgur-Wise'' function.  This allows us to
apply the constraint on $F$ to the particular combination of form
factors actually measured, rather than deriving constraints for each
form factor separately.  For large $b$-mass, the Isgur-Wise function
is given by the
form factor $F$, $F(\w) =  \xi(\w)$, where $\w=v\cdot v'=p_B\cdot
p_D/m_Bm_D$.   
There are  short-distance matching and running
corrections\ref\match{{\HF A. F.  Falk, et al}, \np{B343}{1990}{1}\semi
{\HF A. F.  Falk and B. Grinstein}, \pl{B247}{1990}{406}\semi
{\HF A. F. Falk and B. Grinstein}, \pl{B249}{1990}{314}.}\ to the
relation between $ \xi(\w)$ and the $\bar B \to D^*$ form factors.
For example, for the vector current form factor $g$ one may write
$g(z)=\eta_D F(z)$. This relation generally holds to order
$1/M$, but at threshold it holds to 
order~$1/M^2$\ref\luke{{\HF M. E. Luke}, \pl{B252}{1990}{447}.}. We take
$\eta_D=0.985$. 

\subsec{Maximum Likelihood Fit}

        Once the essential physics of QCD is incorporated into the
calculation via Eq.~\parametriztn, the maximum likelihood fit is
simply an ordinary chi-squared minimization with parameters
$|V_{cb}|$, $a_1$ and $a_2$.  We normalize input data to a $B$
lifetime\ref\blife{{\HF W. Venus}, in {\it Lepton and Photon
Interactions, XVI International Symposium}, edited by {\HF Persis
Drell} and {\HF David Rubin} (AIP Press, New York) 1994.} of $\tau_B =
1.61$ ps. Also, we rescale the  ARGUS data to reflect
a revised branching fraction for $D^0\to K^-\pi^+$.

     	Table 1 shows the central values and $68\%$ confidence levels
for $|V_{cb}|$, $a_1$, and $a_2$ from the various experiments. That $a_2$ 
saturates the bound  $|a_2|\le0.55$ is not significant because its
variance is large.
   	Figure 1 shows the product of the best fit form factors with
$|V_{cb}|$, superimposed with experimental data.  At 90\% confidence
level, $a_1$ and $a_2$ are consistent with zero, suggesting the
dispersion relation may be saturated entirely by higher states.

\vbox{
\hfil\vbox{\offinterlineskip
\hrule
\halign{&\vrule#&\strut\quad\hfil$#$\quad\cr
height2pt&\omit&&\omit&&\omit&&\omit&\cr
&|V_{cb}|\cdot 10^3&&a_1\hfil&&a_2\hfil&&\rm{Expt.}\hfil&\cr
height3pt&\omit&&\omit&&\omit&&\omit&\cr
\noalign{\hrule}
height2pt&\omit&&\omit&&\omit&&\omit&\cr
&35.7_{-2.8}^{+4.2}&&0.00_{-0.07}^{+0.02}&&-0.55_{-0.0}^{+1.1}&&
\rm{CLEO}\cleo &\cr 
&45.4_{-10.7}^{+7.5}&&-0.11_{-0.03}^{+0.10}
&&0.55_{-1.1}^{+0.0}&& \rm{ARGUS}\arg &\cr
&32.2_{-5.89}^{+4.58}&&0.00_{-0.04}^{+0.11}&&0.45_{-1.1}^{+0.1}&&
\rm{ALEPH}\aleph &\cr }
\hrule}
~~~~~~~~~~~~~\hfill}
\medskip
\INSERTCAP{1}{Fit values
for $|V_{cb}|$, $a_1$, and $a_2$ from the various experiments.}

The errors on $|V_{cb}|$ in Table\ 1 are statistical only; the
treatment of systematic errors depends both on our parametrization and
a detailed understanding of the experiment.  The error implicit in the
variation over choices of parametrization, however, is absent.  ARGUS
examined the effect of varying over four possible parametrizations,
which induced a spread of 0.012 in $|V_{cb}|$, and was the major
impediment in using heavy quark symmetry to obtain a model-independent
extraction.

\epsfbox{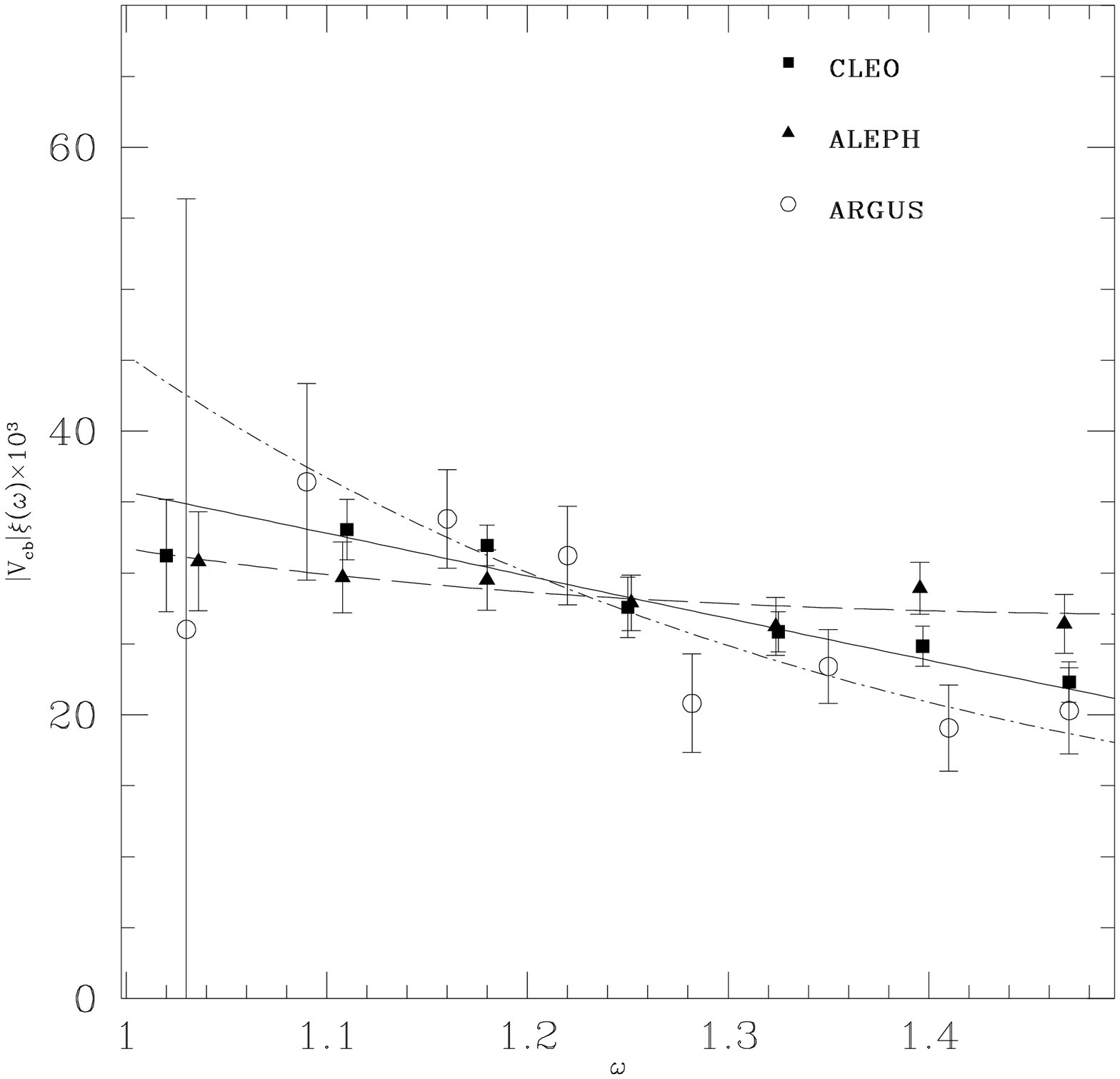}
\vbox{%
{\narrower\noindent%
\multiply\baselineskip by 3%
\divide\baselineskip by 4%
{\rm Figure 1. }{\sl Best fit values for the product of $|V_{cb}|$
with the Isgur-Wise function for CLEO (solid line), ARGUS (dot-dashed
line), and ALEPH (dashed line) data. The data for each experiment,
adjusted for the $B$ lifetime and zero-recoil normalization used in
the text, is superimposed. The ARGUS data has been adjusted to reflect
a revised branching fraction for $D^0\to K^-\pi^+$. \medskip}}}

\newsec{Discussion of the Method}

    	Perturbative and non-perturbative corrections to the
dispersion relation \Idef\ enter our analysis by modifying the value
of the parameter $\rho$ in Eq.~\phidef. The extraction of $|V_{cb}|$
is rather insensitive to such corrections. A change of $\rho$ by
$\pm 10\%$ changes the central value of $|V_{cb}|$ by less than $\pm
0.3\%$.

    	The truncation of Eq.~\basisfn a\ to the first  $N$ terms
introduces an error proportional to ${1\over P(0) \phi(0)}
\sum_{n=N+1}^{\infty} a_{n} z^{n}< 1\%$, a bound valid over the whole
physical region $0\le z\le0.056$.

	The application of the $B \to B$ dispersion relation to $\bar
B \to D^* l \bar \nu$ decays relies on heavy quark symmetry. This is
potentially the largest source of error, of $\CO({1\over M})$. We
estimate such corrections by making a $20\%$ change in the ranges of
$a_1$ and $a_2$, resulting in a $2\%$ shift in the central value of
$|V_{cb}|$. One can derive analogous dispersion relations for each of
the $\bar B \to D^* l
\bar\nu$ form factors, sidestepping the need for heavy quark
symmetry (except at $q^2_{\rm max}$)\ref\uslatest{{\HF C. G. Boyd,
B. Grinstein and R. F. Lebed},  {\sl
Univ.\ of Calif., San Diego Report} No.\ UCSD/PTH~95-11
\href{http://xxx.lanl.gov/abs/hep-ph/9508211}{[hep-ph/9508211]}.}.

	Another heavy quark correction arises at $\CO({1\over M^2})$
in the normalization of the Isgur-Wise function at threshold.  The
normalization of the form factor $g(\w=1)$ has been estimated to be
$g(1)= 0.96$\ref\man{{\HF T. Mannel}, \physrev{D50}{1994}{428}.},
$g(1)= 0.89$\ref\shifman{{\HF M. Shifman, N. Uraltsev, and A.
Vainshtein}, \physrev{D51}{1995}{2271}.}, and $g(1)=
0.93$\ref\falkneubert{{\HF A. F.  Falk and M. Neubert},
\physrev{D47}{1993}{2695} and {\it idem,\/} p.~2982\semi {\HF M.
Neubert}, \pl{B338}{1994}{84}.}. We have included a QCD correction of
$0.985$, so to good approximation, this simply rescales the values of
$|V_{cb}|$ in Table\ 1 by ${0.985 \over g(1)}$.

    	There are other errors in our extraction that are not purely
theoretical. The most pressing of these involve the binning of the
measured rate against $\omega$, smearing of $\omega$ introduced by
boosting from the lab to the center of mass frame, and correlation of
errors.  Randomly varying input values of $\omega$ in our least
squares fit of the CLEO data by $\pm 0.05$ changes the central value
of $|V_{cb}|$ by less than $1\%$.  A more thorough extraction can be
done by the experimental groups themselves, using our parametrization
in their analysis.

\newsec{Other Applications.}

	Our parametrization of form factors applies to other heavy
hadron decays, including $\bar B \to \pi l \bar \nu$, with minor
modifications. In this case the range of the kinematic variable is
larger, $0< z< 0.5$, so more coefficients $a_n$ are needed for
comparable accuracy.  We expect six to eight $a_n$ will be necessary
for accuracy of a few percent over the entire kinematic range,
depending on the form of the actual data.

Heavy quark symmetry relates the form factors for $ B \to \pi l \bar
\nu$ and $D \to\pi l \bar \nu$. The latter is readily measured over
$0< z< 0.3$, and our method then allows a reliable extrapolation to
$0.3\le z<0.5$. This opens up the possibility of reliably extracting
$|V_{ub}|$ from a measurement of ${\rm Br}(B \to \pi l \bar\nu)$.

An extraction of a model-independent lower bound on
$|V_{ub}|$ should be possible since small values of $|V_{ub}|$ tend to
wash out the nontrivial $z$ dependence, while the $a_n{}'s$ cannot
compensate because they are bounded from above.

	 Any model of hadronic form factors must predict a form factor
that is consistent with our parametrization. This is a severe test to
pass, and serves as an effective discriminator for models\ustwo.

	Our parametrization may also be useful in the analysis of $B\to
K^*\gamma$, by relating its amplitude to the form factor for $D\to K^*
e\nu$ extrapolated outside the physical region\ref\iwkstar{{\HF
N. Isgur and M. B. Wise}, \physrev{D42}{1990}{2388}.}.

\newsec{Summary} 

    	The extraction of the CKM mixing parameter $|V_{cb}|$ involves
several types of uncertainties. Typically, these uncertainties are
classified as
\eqn\uncer{
|V_{cb}| = V \pm \{ \it{stat} \} \pm \{ \it{syst} \}
\pm \{ \it{life} \} \pm \{ \it{norm} \} \pm \{
\it{param}
\}
}
where {\it stat} and {\it syst} refer to statistical and systematic
experimental uncertainties, {\it life} refers to uncertainties in the
$B$ lifetime, {\it norm} refers to uncertainty in the value of the
form factor at threshold, and {\it param} refers to uncertainty in the
extrapolation of the measured differential rate to threshold.

   	Not only the central value, but also the statistical
uncertainty depends on the parametrization.  For example, linear fits
to CLEO data yield substantially smaller statistical uncertainties
than quadratic fits. Typically, quoted values correspond to the
parametrization yielding the smallest statistical uncertainty, in
effect throwing some statistical uncertainty into the parametrization
uncertainty, which remains implicit. Clearly, this does not improve
the accuracy with which we know $|V_{cb}|$.

  	We have essentially eliminated the uncertainty
in the choice of parametrization.  This was accomplished in four
stages.  First, we used QCD dispersion relations to constrain the $B
\to B$ elastic form factor.  Second, we derived a general parametrization of
the $B$ elastic form factor that automatically satisfies the dispersion
relation constraint. This expression involved an infinite number of
parameters $a_n$ bounded by $\sum_{n=0}^{\infty} |a_n|^2 \le I$.
Third, we used heavy quark symmetry to relate the $B$ elastic form
factor to $\bar B \to D^* l \bar
\nu$ form factors and fixed the normalization at threshold.  Over the
entire kinematic range relevant to $\bar B \to D^* l \bar \nu$, we
showed that neglecting all but the first two parameters $a_{1}$, $a_{2}$
resulted in at most a $1\%$ deviation in the predicted form factor.
Finally, we made a least squares fit of the differential $\bar B \to
D^* l \bar \nu$ rate to $|V_{cb}|$, $a_1$, and $a_2$.

   	The results of this fit improve on all previous extractions in
one important way: The uncertainty due to the choice of
parametrization is under control.  Our statistical errors are larger
than many quoted values. This does not reflect an inferiority of our
method, but rather quantifies uncertainties that were previously left
implicit.  Our averaged value from CLEO, ARGUS, and ALEPH data is
\eqn\avgv{
|V_{cb}|\times 10^{3}= 35.5^{+2.9}_{-2.5}\, (\it{stat}).}
An estimation of systematic uncertainties requires a detailed
knowledge of the experiments.

   	Our parametrization may be useful for other processes, such as
$\bar B \to \pi l \bar \nu$ and $B\to K^*\gamma$ as well. For $\bar B
\to \pi l \bar \nu$  we expect to be able to extract a model
independent lower bound on $|V_{ub}|$, and a good measure of
$|V_{ub}|$ by use of heavy quark symmetries to infer the form factor
from $D\to \pi l \bar \nu$ by extrapolation. In addition, precision
tests of QCD-predicted form factors are now possible; these should be
useful as checks of QCD models and lattice simulations.

\vskip1.2cm
{\it Acknowledgments}\hfil\break We would like to thank Vivek Sharma,
Hans Paar and Persis Drell for useful discussions.  The research of
one of us (B.G.)  is funded in part by the Alfred P. Sloan
Foundation. This work is supported in part by the Department of Energy
under contract DOE--FG03--90ER40546.

\vfill\eject
\listrefs
\bye